\documentclass[useAMS,usenatbib]{mn2e}

\usepackage{graphicx} 

\title[Location and energetics of ultra-fast outflows]{Evidence for ultra-fast outflows in radio-quiet AGNs: III - location and energetics}
\author[F. Tombesi et al.]{F. Tombesi$^{1,2}$\thanks{E-mail: ftombesi@astro.umd.edu}, M. Cappi$^3$, J.~N. Reeves$^4$, and V. Braito$^{5,6}$\\
$^1$X-ray Astrophysics Laboratory and CRESST, NASA/Goddard Space Flight Center, Greenbelt, MD 20771, USA\\
$^2$Department of Astronomy, University of Maryland, College Park, MD 20742, USA\\
$^3$INAF-IASF Bologna, Via Gobetti 101, I-40129 Bologna, Italy\\
$^4$Astrophysics Group, School of Physical and Geographical Sciences, Keele University, Keele, Staffordshire ST5 5BG, UK\\
$^5$Department of Physics and Astronomy, University of Leicester, University Road, Leicester LE1 7RH, UK\\
$^6$INAF-Osservatorio Astronomico di Brera, via. E. Bianchi 46, I-23807, Merate, Italy}
\begin{document}

\date{Accepted ???. Received ???; in original form ???}

%\pagerange{\pageref{firstpage}--\pageref{lastpage}} \pubyear{2002}

\maketitle

\label{firstpage}

\begin{abstract}

Using the results of a previous X-ray photo-ionization modelling of blue-shifted Fe K absorption lines on a sample of 42 local radio-quiet AGNs observed with \emph{XMM-Newton}, in this letter we estimate the location and energetics of the associated ultra-fast outflows (UFOs). 
Due to significant uncertainties, we are essentially able to place only lower/upper limits. On average, their location is in the interval $\sim$0.0003--0.03pc ($\sim$$10^2$--$10^4$$r_s$) from the central black hole, consistent with what is expected for accretion disk winds/outflows.
The mass outflow rates are constrained between $\sim$0.01--1~$M_{\odot}$~yr$^{-1}$, corresponding to $\ga$5--10\% of the accretion rates. The average lower--upper limits on the mechanical power are log$\dot{E}_K$$\simeq$42.6--44.6~erg~s$^{-1}$. 
However, the minimum possible value of the ratio between the mechanical power and bolometric luminosity is constrained to be comparable or higher than the minimum required by simulations of feedback induced by winds/outflows.
Therefore, this work demonstrates that UFOs are indeed capable to provide a significant contribution to the AGN cosmological feedback, in agreement with theoretical expectations and the recent observation of interactions between AGN outflows and the interstellar medium in several Seyferts galaxies.

\end{abstract}


\section{Introduction}

\begin{table*}
\centering
\begin{minipage}{170mm}
\caption{Location and energetics of the Fe K absorbers.}
\begin{tabular}{@{\hspace{0.3cm}}r@{\hspace{0.3cm}}l@{\hspace{0.3cm}}c@{\hspace{0.3cm}}c@{\hspace{0.3cm}}c@{\hspace{0.3cm}}c@{\hspace{0.3cm}}c@{\hspace{0.3cm}}c@{\hspace{0.3cm}}c@{\hspace{0.3cm}}c@{\hspace{0.3cm}}c@{\hspace{0.3cm}}}
\hline
 & Source & log$M_{BH}$ & XMM Obs & log$L^*$ & log$r_{min}$ & log$r_{max}$ & log$\dot{M}_{out}^{min}$ & log$\dot{M}_{out}^{max}$ & log$\dot{E}_{K}^{min}$ & log$\dot{E}_{K}^{max}$\\
 & & ($M_{\odot}$) & & (erg~s$^{-1}$) & (cm) & (cm) & (g~s$^{-1}$) & (g~s$^{-1}$) & (erg~s$^{-1}$) & (erg~s$^{-1}$)\\
\hline\\[-11.5pt]
\multicolumn{11}{c}{UFOs}\\[-3.5pt]
\hline
1 & NGC~4151 & $7.1\pm0.2^1$ & 0402660201 & 42.5/42.9 & $14.6\pm0.2$ & $<15.8$ & $>23.2$ & $24.4\pm0.5$ & $>41.9$ & $43.1\pm0.5$\\
2 & IC4329A  & $8.1\pm0.2^2$ & 0147440101 & 43.7/44.1 & $15.6\pm0.2$ & $<16.5$ & $>24.2$ & $25.0\pm0.9$ & $>42.8$ & $43.6\pm0.9$\\
3 & Mrk~509  & $8.1\pm0.1^1$ & 0130720101 & 43.9/44.2 & $15.1\pm0.1$ & $<16.3$ & $>24.4$ & $25.7\pm0.6$ & $>43.5$ & $44.8\pm0.6$\\
4 &          &               & 0306090201 & 44.0/44.4 & $15.3\pm0.1$ & $<16.6$ & $>24.5$ & $25.8\pm1.0$ & $>43.4$ & $44.7\pm1.0$\\
5 &          &               & 0306090401 & 44.0/44.4 & $14.9\pm0.1$ & $<18.1$ & $>23.5$ & $26.8\pm1.5$ & $>42.8$ & $46.1\pm1.5$\\
6 & Ark~120  & $8.2\pm0.1^1$ & 0147190101 & 44.0/44.5 & $14.8\pm0.1$ & $<17.9$ & $>23.5$ & $26.7\pm1.3$ & $>43.1$ & $46.2\pm1.3$\\
7 & Mrk~79   & $7.7\pm0.1^1$ & 0400070201 & 43.4/43.9 & $15.3\pm0.1$ & $16.5\pm0.4$ & $24.7\pm0.3$ & $26.0\pm0.2$ & $43.3\pm0.3$ & $44.6\pm0.2$\\
8 & NGC~4051 & $6.3\pm0.4^4$ & 0109141401 & 41.5/42.3 & $14.7\pm0.7$ & $<15.9$ & $>22.5$ & $23.8\pm1.6$ & $>40.3$ & $41.6\pm1.7$\\
9 &          &               & 0157560101 & 41.0/42.0 & $13.2\pm0.2$ & $16.2\pm0.2$ & $22.5\pm0.2$ & $25.5\pm0.2$ & $41.8\pm0.2$ & $44.8\pm0.2$\\
10 & Mrk~766 & $6.1\pm0.4^4$ & 0304030301 & 42.6/43.2 & $13.8\pm0.4$ & $17.2\pm0.5$ & $22.3\pm0.4$ & $25.7\pm0.5$ & $40.8\pm0.4$ & $44.2\pm0.5$\\
11 &         &               & 0304030501 & 42.8/43.4 & $13.7\pm0.4$ & $16.1\pm0.2$ & $22.9\pm0.4$ & $25.3\pm0.1$ & $41.4\pm0.4$ & $43.8\pm0.1$\\
12 & Mrk~841 & $7.8\pm0.5^5$ & 0205340401 & 43.5/43.9 & $15.8\pm0.6$ & $<18.0$ & $>23.8$ & $26.0\pm1.2$ & $>41.9$ & $44.1\pm1.2$\\
13 & 1H0419-577 & $8.6\pm0.5^3$ & 0148000201 & 44.3/44.6 & $16.3\pm0.5$ & $17.9\pm0.7$ & $25.5\pm0.7$ & $27.1\pm0.5$ & $43.9\pm0.7$ & $45.5\pm0.5$\\
14 & Mrk~290 & $7.7\pm0.5^5$ & 0400360601 & 43.2/43.6 & $14.8\pm0.5$ & $16.7\pm1.3$ & $24.3\pm0.9$ & $26.2\pm1.2$ & $43.4\pm0.9$ & $45.3\pm1.2$\\
15 & Mrk~205 & $8.6\pm1.0^6$ & 0124110101 & 43.8/44.2 & $16.1\pm1.0$ & $<16.2$ & $>25.6$ & $25.6\pm0.6$ & $>44.1$ & $44.3\pm0.6$\\
16 & PG~1211$+$143 & $8.2\pm0.2^1$ & 0112610101 & 43.7/44.3 & $15.3\pm0.2$ & $18.5\pm0.1$ & $24.7\pm0.2$ & $27.9\pm0.1$ & $43.7\pm0.2$ & $46.9\pm0.1$\\
17 & MCG-5-23-16 & $7.6\pm1.0^6$ & 0302850201 & 43.1/43.5 & $15.0\pm1.0$ & $16.6\pm0.1$ & $23.9\pm1.0$ & $25.5\pm0.1$ & $42.7\pm1.0$ & $44.3\pm0.2$\\
18 & NGC~4507 & $6.4\pm0.5^5$ & 0006220201 & 43.1/43.4 & $13.3\pm0.5$ & $<16.9$ & $>21.9$ & $25.4\pm1.1$ & $>41.2$ & $44.6\pm1.1$\\
19 & NGC~7582 & $7.1\pm1.0^6$ & 0112310201 & 41.6/42.0 & $13.7\pm1.0$ & $15.2\pm0.3$ & $23.8\pm1.0$ & $25.3\pm0.1$ & $43.4\pm1.1$ & $44.9\pm0.1$\\
\hline\\[-11.5pt]
\multicolumn{11}{c}{non-UFOs}\\[-3.5pt]
\hline
20 & NGC~3783 & $7.5\pm0.1^1$ & 0112210101 & 43.1/43.6 & $17.0\pm0.4$ & $19.1\pm0.2$ & $24.7\pm0.4$ & $26.7\pm0.2$ & $41.3\pm0.5$ & $43.4\pm0.4$\\
21 &          &               & 0112210201 & 43.0/43.4 & $>17.3$ & $18.1\pm0.1$ & $>24.8$ & $<25.7$ & $>41.1$ & $<42.0$\\
22 &          &               & 0112210501 & 43.1/43.5 & $>17.3$ & $18.1\pm0.1$ & $>24.8$ & $<25.6$ & $>41.1$ & $<42.0$\\
23 & NGC~3516 & $7.2\pm0.2^7$ & 0401210401 & 43.0/43.8 & $17.1\pm0.3$ & $17.1\pm0.2$ & $24.8\pm0.4$ & $24.8\pm0.2$ & $41.0\pm0.5$ & $41.0\pm0.3$\\
24 &          &               & 0401210501 & 43.0/43.7 & $16.8\pm0.3$ & $16.6\pm0.1$ & $24.9\pm0.3$ & $24.8\pm0.1$ & $41.3\pm0.4$ & $41.3\pm0.2$\\
25 &          &               & 0401210601 & 42.9/43.6 & $16.6\pm0.2$ & $16.7\pm0.2$ & $24.7\pm0.3$ & $24.9\pm0.1$ & $41.4\pm0.3$ & $41.6\pm0.2$\\
26 &          &               & 0401211001 & 43.0/43.7 & $16.4\pm0.3$ & $16.7\pm0.2$ & $24.6\pm0.4$ & $24.9\pm0.1$ & $41.4\pm0.4$ & $41.8\pm0.2$\\
27 & Mrk~279  & $7.5\pm0.2^1$ & 0302480501 & 43.7/44.1 & $>17.3$ & $17.9\pm0.7$ & $>24.9$ & $<25.5$ & $>41.2$ & $<41.8$\\
28 & ESO~323-G77 & $7.4\pm0.5^5$ & 0300240501 & 43.0/44.0 & $16.7\pm0.6$ & $17.0\pm0.5$ & $25.3\pm0.7$ & $25.6\pm0.4$ & $42.1\pm0.7$ & $42.4\pm0.5$\\
\hline
\end{tabular}
$^*$ 2--10~keV luminosity $L_{2-10}$ over ionizing luminosity $L_{ion}$; $^1$ Peterson et al.~(2004); $^2$ Markowitz et al.~(2009); $^3$ Bian \& Zhao (2003); $^4$ Bentz et al.~(2009); $^5$ Wang \& Zhang (2007); $^6$ Wandel \& Mushotzky (1986); $^7$ Onken et al.~(2003) .
\end{minipage}
\end{table*}

Blueshifted Fe K-shell absorption lines have been detected in recent years in the X-ray spectra of several radio-quiet AGNs (Chartas et al.~2002, 2003; Pounds et al.~2003; Markowitz et al.~2006; Braito et al.~2007; Cappi et al.~2009; Reeves et al.~2009; Giustini et al.~2011). These findings are important because they suggest the presence of massive and highly ionized absorbers outflowing from their nuclei with mildly-relativistic velocities. They are possibly connected with accretion disc winds/outflows (King \& Pounds 2003; Proga \& Kallman 2004; Ohsuga et al.~2009; Sim et al.~2010) or the base of a possible weak jet (e.g., Ghisellini et al.~2004).
In particular, a uniform and systematic search for blueshifted Fe K absorption lines in a sample of 42 local ($z$$\le$0.1) radio-quiet AGNs observed with \emph{XMM-Newton} was performed by Tombesi et al.~(2010a, hereafter paper I). This allowed the authors to assess their global significance and derive a detection fraction of $\ga$40\%. 
In order to have a clear distinction with the classical soft X-ray warm absorbers, in paper I we defined Ultra-fast Outflows (UFOs) as those highly ionized Fe K absorbers with blueshifted velocity $\ge$10,000~km/s. In fact, the warm absorbers are usually less ionized, have outflow velocities in the range $\sim$100--1000~km/s and may possibly have a different physical origin (Blustin et al.~2005; McKernan et al.~2007). In the following we refer to the Fe K absorbers with outflow velocity $<$10,000~km/s as non-UFOs.
Then, Tombesi et al.~(2011a, hereafter paper II) performed a photo-ionization modelling and derived the distribution of the main physical parameters. The outflow velocity is mildly-relativistic, in the range $\sim$0.03--0.3c, with a peak and mean value at $\sim$0.14c. The ionization is very high, in the range log$\xi$$\sim$3--6~erg~s$^{-1}$~cm, with a mean value of $\sim$4.2~erg~s$^{-1}$~cm. The column densities are also large, in the interval $N_H$$\sim$$10^{22}$--$10^{24}$~cm$^{-2}$, with a mean value of $\sim$$10^{23}$~cm$^{-2}$. 
It is important to note that Tombesi et al.~(2010b, 2011b) detected the presence of UFOs also in a small sample of radio-loud AGNs observed with \emph{Suzaku}. 

In this letter we will constrain the distance of UFOs from the central super-massive black hole (SMBH) and we will also quantify their energetics and mass content, which are crucial for the understanding of their contribution to the overall energetic budget of AGNs and possible feedback impact on the surrounding environment. The analysis of the possible correlations among the parameters and a comparison with the soft X-ray warm absorbers is postponed to a successive paper IV of this series.

\section[]{Location and energetics}

We base our estimates using the outflow velocity, ionization parameter and column density of the Fe K absorbers reported in Table~3 of paper II.   
The sources and relative \emph{XMM-Newton} observations are reported in Table~1. There, we also list the estimated SMBH masses and the absorption corrected X-ray luminosities calculated in the 2--10~keV and 1--1000~Ryd (1~Ryd=13.6~eV; see column 5). 

An estimate of the maximum distance from the central source can be derived from the definition of the ionization parameter $\xi=L_{ion}/nr^2$ (Tarter et al.~1969). For compact absorbers we obtain $r \le r_{max}=L_{ion}/\xi N_H$. Instead, 
an estimate of the minimum distance can be derived from the radius at which the observed velocity corresponds to the escape velocity, $r \ge r_{min}=2 G M_{BH}/ v_{out}^{2}$. The derived values and errors are reported in Table~1 and Fig.~1. The average location of UFOs and non-UFOs is between $\sim$0.0003--0.03pc ($\sim$$10^2$--$10^4$$r_s$, $r_s$$=$$2GM_{BH}/c^2$) and $\sim$0.03--0.3pc ($\sim$$10^4$--$10^5$$r_s$), respectively. Both of these ranges are within, or comparable to, the typical location of the soft X-ray warm absorbers, at $\sim$pc scales (Blustin et al.~2005; McKernan et al.~2007). Therefore, this strongly suggests a direct identification with accretion disc winds/outflows. It is also important to note that there is a continuity between the two intervals, with the UFOs systematically closer in. The observed spectral variability, even on time-scales of $\sim$days in some cases (e.g., Braito et al.~2007; Cappi et al.~2009; Tombesi et al.~2011b; paper I), is also consistent with the assumption of compact absorbers and the location being close to the SMBH. This also suggests that they are probably intermittent and/or clumpy.

  \begin{figure}
  \centering
   \includegraphics[width=5.5cm,height=7.5cm,angle=-90]{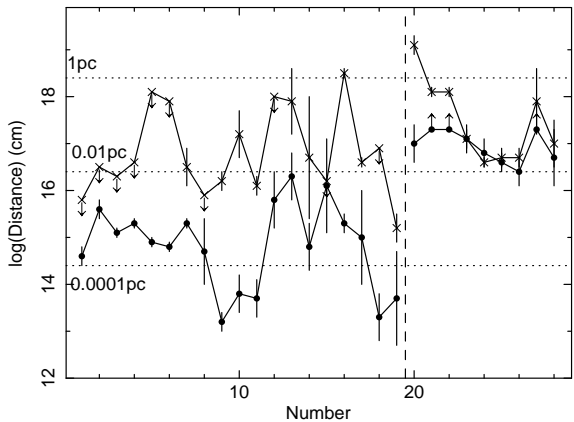}
   \caption{Lower (filled circles) and upper limits (crosses) on the distance of the Fe K absorbers from the central SMBH. The vertical line separates the UFOs (left) and non-UFOs (right).}
    \end{figure}

We use the expression for the mass outflow rate derived by Krongold et al.~(2007), which is more appropriate for a biconical wind-like geometry instead of a simple spherical one: $\dot{M}_{out}=0.8 \pi m_p N_H v_{out} r f(\delta, \phi)$. $f(\delta, \phi)$ is a function that depends on the angle between the line of sight to the central source and the accretion disc plane, $\delta$, and the angle formed by the wind with the accretion disc, $\phi$ (see Fig. 12 of Krongold et al.~2007). 
For a vertical disc wind ($\phi$$=$$\pi/2$) and an average line-of-sight angle $\delta$$=$$30^{\circ}$ for the Seyferts considered here, $f(\delta, \phi)$$\simeq$1.5. This mass outflow rate formula has also the important advantage of not relying on the estimate of the covering and filling factors. This is due to the fact that it takes into account only the net observed thickness of the gas, allowing for clumping in the flow. Thus, there is not the need to include a linear (or volume) filling factor, since we are interested in estimating the net flow of mass, starting from the observed column density and velocity. Moreover, the covering factor is implicitly taken into account by the function $f(\delta, \phi)$ when calculating the area filled by the gas, constrained between the inner and outer conical surfaces.
The assumptions are that the thickness of the wind between the two conical surfaces is constant with $\delta$ and that this is much smaller than the distance to the source. Full details on the derivation of this formula can be found in the Appendix 2 of Krongold et al.~(2007).
However, it is important to note that we obtain equivalent results including a clumpiness factor of $\Delta R/R$ along the line of sight in the spherical approximation case (Tombesi et al.~2010b, 2011b) and using a covering fraction $C$$\simeq$$0.2 f(\delta, \phi)$$\simeq$0.4, which is consistent with the value derived observationally from the detection fraction of UFOs in paper I and II.  
Using the lower/upper limits on the distance we can thus estimate the lower/upper limits on the mass outflow rate and relative errors, see Table~1 and Fig.~2. The average values are in the range $\sim$0.01--1~$M_{\odot}$~yr$^{-1}$ for the UFOs and $\sim$0.1--0.5~$M_{\odot}$~yr$^{-1}$ for the non-UFOs, respectively. They are consistent with each other.

The kinetic or mechanical power of the outflows can be estimated as $\dot{E}_K = \frac{1}{2} \dot{M}_{out} v_{out}^2$. The lower/upper limits and relative errors are reported in Table~1 and Fig.~3. 
The average values for UFOs and non-UFOs are log$\dot{E}_K$$\simeq$42.6--44.6~erg~s$^{-1}$ and log$\dot{E}_K$$\simeq$41.3--42~erg~s$^{-1}$, respectively. This is comparable to the X-ray ionizing luminosity $L_{ion}$ and, again, there is a continuity between the two intervals, with UFOs having systematically higher values. Theoretical models and simulations show that the mechanical power needed by accretion disc winds/outflows in order to have a significant feedback impact on the surrounding environment is typically about $\sim$5\% of the bolometric luminosity (Di Matteo et al.~2005; King 2010; Ostriker et al.~2010; DeBuhr et al.~2011). However, a recent work by Hopkins \& Elvis (2010) demonstrated that the minimum ratio required is actually only $\sim$0.5\%. Using the lower limits on the mechanical power and the upper limit on the bolometric correction of $K_{2-10}$$<$100 (see \S3), we can derive an average lower limit of $\dot{E}_K/L_{bol}$$>$0.3\% for the UFOs. We stress that this is the minimum possible value. In fact, given the uncertainty on the bolometric correction and using the average upper limits on $\dot{E}_K$, we obtain a maximum value that can potentially be comparable to $L_{bol}$.
Therefore, despite the significant uncertainties, we find that this ratio is comparable or higher than the minimum value required to imprint a significant feedback. The relative value for the non-UFOs is instead lower, $\dot{E}_K/L_{bol}$$\sim$0.02--0.8\%, but still possibly capable to generate at least a weak feedback.

  \begin{figure}
  \centering
   \includegraphics[width=5.5cm,height=7.5cm,angle=-90]{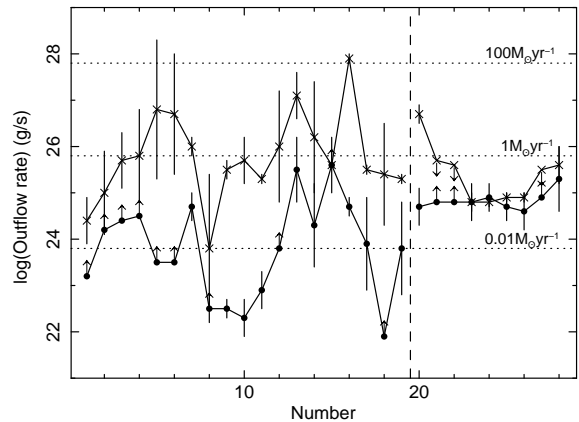}
   \caption{Lower (filled circles) and upper limits (crosses) on the mass outflow rate of the Fe K absorbers. The vertical line separates the UFOs (left) and non-UFOs (right).}
    \end{figure}

As previously derived, the mass outflow rate can be significant, even of the order of $\sim$1~$M_{\odot}$~yr$^{-1}$ or higher. It is then interesting to know how this compares to the accretion rate, $\dot{M}_{acc}$$=$$L_{bol}/\eta c^2$. To quantify this we need to know the radiative efficiency $\eta$. As discussed in \S3, this is not well determined for each source and the uncertainties on $\dot{M}_{acc}$ can be significant. Therefore, considering an upper limit $K_{2-10}$$<$100 and a lower limit $\eta$$\ga$0.05, we estimate that $\dot{M}_{out}/\dot{M}_{acc}$$\ga$5--10\% for both UFOs and non-UFOs. However, given the significant uncertainties, the mass outflow rate could potentially exceed the accretion rate in some cases.  
Finally, due to the large uncertainties on the parameters in Table 1, we can not significantly constrain any variability of the outflow properties for the five sources with multiple observations.

\section[]{Error analysis}

In the calculation of the parameters reported in Table~1 we took into account the propagation of errors on the ionization parameter, column density, outflow velocity and SMBH mass. Here we discuss in more detail the possible sources of systematic uncertainty.

In order to limit the uncertainty on the slope of the ionization continuum, in paper II we estimated that the average SED of the sources corresponds to a $\Gamma$$\simeq$2 power-law with high energy cut-off at E$\simeq$100~keV in the input energy range for the photo-ionization code {\sc Xstar}. Observationally, this is in agreement with the result of a systematic spectral analysis of Seyfert 1s observed with \emph{BeppoSAX} in the 2--100~keV performed by Dadina (2008), who derived an average $\Gamma \simeq 1.9$ and cut-off at E$\sim$200~keV. 
Even if we limited our analysis in the 4--10~keV, from paper I we can estimate an average $\Gamma$$\sim$1.8 and a scatter of $\sim$0.2. This is consistent with Dadina (2008) and the slightly flatter $\Gamma$ is probably due to an emerging weak reflection component. If we consider this typical scatter, we derive that the possible uncertainty on the slope of the ionizing continuum may induce a maximum systematic error of 0.4~dex on the ionization parameter.

  \begin{figure}
  \centering
   \includegraphics[width=5.5cm,height=7.5cm,angle=-90]{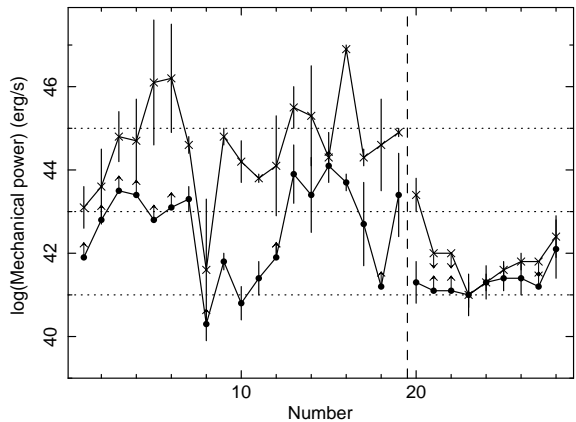}
   \caption{Lower (filled circles) and upper limits (crosses) on the mechanical power of the Fe K absorbers. The vertical line separates the UFOs (left) and non-UFOs (right).}
    \end{figure}

We note that Standard Solar abundances from Asplund et al.~(2009) were assumed in paper II. If the iron abundance is allowed to be $\pm$2 times Solar, the resultant values are still consistent within the 1$\sigma$ errors, with a typical difference $\la$0.2~dex. 
We point out that when performing the photoionization modelling of the absorption lines in paper II, it was not possible to clearly distinguish their identification as due predominantly to Fe~XXV or Fe~XXVI in 6/28 observations. In these cases we obtained two solutions with similar reduced $\chi^2$ but different values of the ionization parameter, column density and velocity. However, this uncertainty was taken into account when calculating the relative errors on the parameters reported in Table~3 of paper II. 
Regarding the SMBH masses, the possible systematic uncertainty for those derived using reverberation mapping techniques is $<$0.5~dex (e.g., Peterson et al.~2004). 
We note that the expression for the mass outflow rate used in \S2 has a possible systematic source of uncertainty from the factor $f(\delta, \phi)$. For all reasonable angles ($\delta$$>$$20^{\circ}$ and $\phi$$>$$45^{\circ}$) this is of the order of unity, with a maximum variation of $\sim$0.3~dex (see Krongold et al.~2007).

The estimate of the bolometric luminosity and radiative efficiency for each source would require a detailed modelling of the SEDs, which is beyond the scope of the present letter. One way to overcome this is using the 2--10~keV luminosity as a proxy and apply a bolometric correction, $L_{bol}$$=$$K_{2-10}$$L_{2-10}$~erg~s$^{-1}$. From the SEDs of the sources analysed in paper II we derive a rough average estimate of $K_{2-10}$$\sim$30. However, it has been reported that there could be a significant scatter of this value in the maximum range of $K_{2-10}$$\simeq$10--100 (Vasudevan \& Fabian 2009; Lusso et al.~2010; Nemmen \& Brotherton 2010). Thus, this translates in a maximum error of $\la$1.4~dex in $\dot{E}_{K}/L_{bol}$.  
The radiative efficiency $\eta$ is also not well known for each source. 
Theoretically, this is in the range $\sim$0.05--0.3, for a non- or maximally rotating black hole (Novikov \& Thorne 1973). Observationally, its average is typically derived using the integrated background luminosity of AGNs and the Soltan argument, obtaining a value of $\eta$$\simeq$0.1 (Soltan 1982; Elvis et al.~2002). Few attempts have been made applying also a detailed source by source analysis. For instance, Davis \& Laor (2011) obtained an average value of log$\eta$$=$$-1.05\pm0.52$. Considering this, we expect a maximum error on the accretion rate of $\sim$1~dex and $\sim$1.5~dex on the ratio $\dot{M}_{out}/\dot{M}_{acc}$.

\section[]{Discussion and conclusions}

In this letter we estimate the location, mass outflow rate and mechanical power of highly ionized Fe K absorbers detected in a large sample of Seyfert galaxies observed with \emph{XMM-Newton}. 
Their parameters show a continuity between those classified as UFOs and non-UFOs (see \S2), with the latter occupying the lower end of the parameter space and suggesting a possible common physical origin.
Indeed, they are directly consistent with an identification as accretion disc winds/outflows, both having velocities higher than most warm absorbers. Intriguing, they might possibly be related also to the radio jet activity (Tombesi et al.~2010b, 2011b). Considering the most pessimistic scenario, we are still able to confirm that the mechanical power of UFOs is indeed sufficient to exert a significant feedback impact on the surrounding environment.  

The cosmological feedback from AGN outflows/jets has been demonstrated to influence the bulge star formation and SMBH growth and possibly also to contribute to the establishment of the observed SMBH-host galaxy relations, such as the $M_{BH}$--$\sigma$ (Di Matteo et al.~2005; King 2010; Ostriker et al.~2010; DeBuhr et al.~2011; Hopkins \& Elvis 2010).
Similar and possibly even more massive and/or energetic outflows might have influenced also the formation of structures and galaxy evolution through feedback at higher redshifts, close to the peak of the quasar activity at $z \sim 2$ (Silk \& Rees 1998; Scannapieco \& Oh 2004; Hopkins et al.~2006).
Simulations of AGN outflows with characteristics equivalent to UFOs have also been independently demonstrated to be able to significantly interact not only with the interstellar medium of the host galaxy but possibly also with the intergalactic medium. They can provide a significant contribution to the quenching of cooling flows and the inflation of bubbles/cavities in the intergalactic medium in both galaxy clusters (e.g., Sternberg et al.~2007; Gaspari et al.~2011a) and especially groups (e.g., Gaspari et al.~2011b).
The UFOs, and AGN outflows in general, might actually provide a feedback impact comparable or even greater than that from jets. 
In fact, the UFOs are likely more massive than jets. They are mildly-relativistic and have somewhat wide angles, therefore possibly exerting a higher impact on the surrounding host galaxy environment compared to the highly collimated relativistic jets, which might actually drill out of the galaxy and have a dominant effect only in the outside. UFOs are energetic, with a mechanical power comparable to that of jets (Tombesi et al.~2010b, 2011b). Moreover, UFOs have been found in $\ga$40\% of local radio-quiet AGNs (papers I and II) and may possibly have a more widespread feedback influence with respect to the less common radio-loud sources with powerful jets. Finally, accretion disc outflows have been found also in radio-loud AGNs (Tombesi et al.~2010b, 2011b) and therefore their feedback effect might actually be concomitant with that from jets.
 
Observationally, we note that direct evidence for AGN feedback activity driven by outflows/jets is recently emerging also for Seyfert galaxies, with the detection of bubbles, shocks and jet/cloud interaction from $\sim$pc up to $\sim$kpc scales (e.g., NGC~4151, Wang et al.~2010; NGC~4051, Pounds \& Vaughan 2011; both part of our sample and with detected UFOs).
In conclusion, there is now plenty of theoretical and observational evidence that AGN feedback through outflows have the possibility to tie together the densest objects at the center of galaxies with the most diffuse regions of intergalactic gas, impacting all intermediate structures. In this regard, this work shows that UFOs provide another important observational piece for the solution of this puzzle.
Significant improvements are expected from the higher effective area and energy resolution in the Fe K band offered by the micro-calorimeters on board \emph{Astro-H} and especially the proposed ESA mission \emph{Athena}.

\section*{Acknowledgments}

FT particularly thank G.~G.~C. Palumbo and R.~M. Sambruna for their support. FT thank R.~F. Mushotzky, C.~S. Reynolds and M. Dadina for the useful discussions. MC acknowledge support from ASI under contract INAF/ASI I/009/10/0. The authors thank the referee for suggestions that led to improvements in the paper.

\end{document}